\documentclass[]{interact}
\usepackage{graphicx}
\usepackage{float}
\usepackage[ruled, vlined, linesnumbered]{algorithm2e}
\usepackage{extraplaceins}
\usepackage{epstopdf}
\usepackage[caption=false]{subfig}

\usepackage[numbers,sort&compress]{natbib}
\bibpunct[, ]{[}{]}{,}{n}{,}{,}
\renewcommand\bibfont{\fontsize{10}{12}\selectfont}

\theoremstyle{plain}
\newtheorem{theorem}{Theorem}[section]
\newtheorem{lemma}[theorem]{Lemma}
\newtheorem{corollary}[theorem]{Corollary}

\theoremstyle{definition}

\theoremstyle{remark}

\begin{document}

\articletype{}

\title{Efficient many-to-many matching of points with demands in one dimension}

\author{
\name{Fatemeh Rajabi-Alni\textsuperscript{a}\thanks{F. Rajabi-Alni. Email: rajabi\_fatemeh@mail.iust.ac.ir; fatemehrajabialni@yahoo.com}  and Behrouz Minaei-Bidgoli\textsuperscript{a}\thanks{B. Minaei-Bidgoli. Email: b\_minaei@iust.ac.ir}
}
\affil{\textsuperscript{a}School of Computer Engineering, Iran University of Science and Technology, Tehran, Iran
}
}

\maketitle

\begin{abstract}
Given two point sets $S$ and $T$, the minimum-cost \textit{many-to-many matching with demands} (MMD) problem is the problem of finding a minimum-cost many-to-many matching between $S$ and $T$ such that each point of $S$ (respectively $T$) is matched to at least a given number of the points of $T$ (respectively $S$). We propose the first $O\left(n^2\right)$-time algorithm for computing a \textit{one dimensional MMD} (OMMD) of minimum cost between $S$ and $T$, where $\left|S\right|+\left|T\right|=n$. In an OMMD problem, the input point sets $S$ and $T$ lie on the real line and the cost of matching a point to another point equals the Euclidean distance between the two points. We also study a generalized version of the MMD problem, the \textit{many-to-many matching with demands and capacities} (MMDC) problem, that in which each point has a limited capacity in addition to a demand. We give the first $O(n^2)$-time algorithm for the minimum-cost \textit{one dimensional MMDC} (OMMDC) problem.
\end{abstract}

\begin{keywords}
Many-to-many point matching; One dimensional point-matching; Demands; Capacities

\end{keywords}


\section{Introduction}
\label{intro}

Suppose we are given two point sets $S$ and $T$ with $|S|+ |T|=n$, a \textit{many-to-many matching} (MM) between $S$ and $T$ assigns each point of one set to one or more points of the other set \citep{ColanDamian}. The minimum-cost MM problem has been solved using the Hungarian method in $O(n^3)$ time \cite{Eiter}. Colannino et al. \cite{ColanDamian} presented an $O(n \log {n})$-time dynamic programming solution for finding an MM of minimum-cost between two sets on the real line. For more discussion on the MM, see \cite{Imanparast,Rajabi-Alni3}.

A general case of the MM problem is the \textit{limited capacity many-to-many matching} (LCMM) problem where each point has a capacity, i.e. each point can be matched to at most a given number of the points. A special case of the LCMM problem, the \textit{one dimensional LCMM} (OLCMM) problem, is that in which both $S$ and $T$ lie on the real line. In \cite{Rajabi-Alni}, an $O(n)$-time efficient algorithm was proposed for the minimum-cost OLCMM.

In this paper, we consider another generalization of the MM problem, where each point has a \textit{demand}, that is each point of one set must be matched to at least a given number of the points of the other set. Let $S=\{s_1,s_2,\dots,s_y\}$ and $T=\{t_1,t_2,\dots,t_z\}$. We denote the demand sets of $S$ and $T$ by $D_S=\{\alpha_1,\alpha_2,\dots,\alpha_y\}$ and $D_T=\{\beta_1,\beta_2,\dots,\beta_z\}$, respectively. In a \textit{many-to-many matching with demand} (MMD), each point $s_i \in S$ must be matched to at least $\alpha_i$ points in $T$ and each point $t_j \in T$ must be matched to at least $\beta_j$ points in $S$. We study the \textit{one dimensional MMD} (OMMD), where $S$ and $T$ lie on the real line and the cost of matching $s_i\in S$ to $t_j\in T$ equals their distance on the line for $1\leq i\leq y$ and $1\leq j\leq z$. We propose an $O(n^2)$ algorithm for finding a minimum-cost OMMD between $S$ and $T$. We also give an $O(n^2)$ time algorithm for a more general version of the OMMD problem, the \textit{one dimensional many-to-many matching with demands and capacities} (OMMDC) problem, where each point has a demand and a capacity.


An important motivation for the study of the OMMD and OMMDC problems is their application in wireless networks. For example, consider a wireless network deployed along a one-dimensional line in applications such as environmental boundary monitoring and target tracking, or performing border patrol, or vehicular ad hoc network on highways \citep{Susca,Frasca,Li,Hu,Aleksandrov}. As an example for the MMD problem, consider the target coverage problem where each target should be monitored by at least a given number of sensors \cite{Yarinezhad2020}. Another important application of the MM problem is data matching \cite{BLUMENTHAL2022202,Fredrickson}.


\section{Preliminaries}
\label{PreliminSect}
\begin{figure}[h]

\hspace{1cm}
\vspace{0.1cm}
\resizebox{0.6\width}{!}{%

  \includegraphics{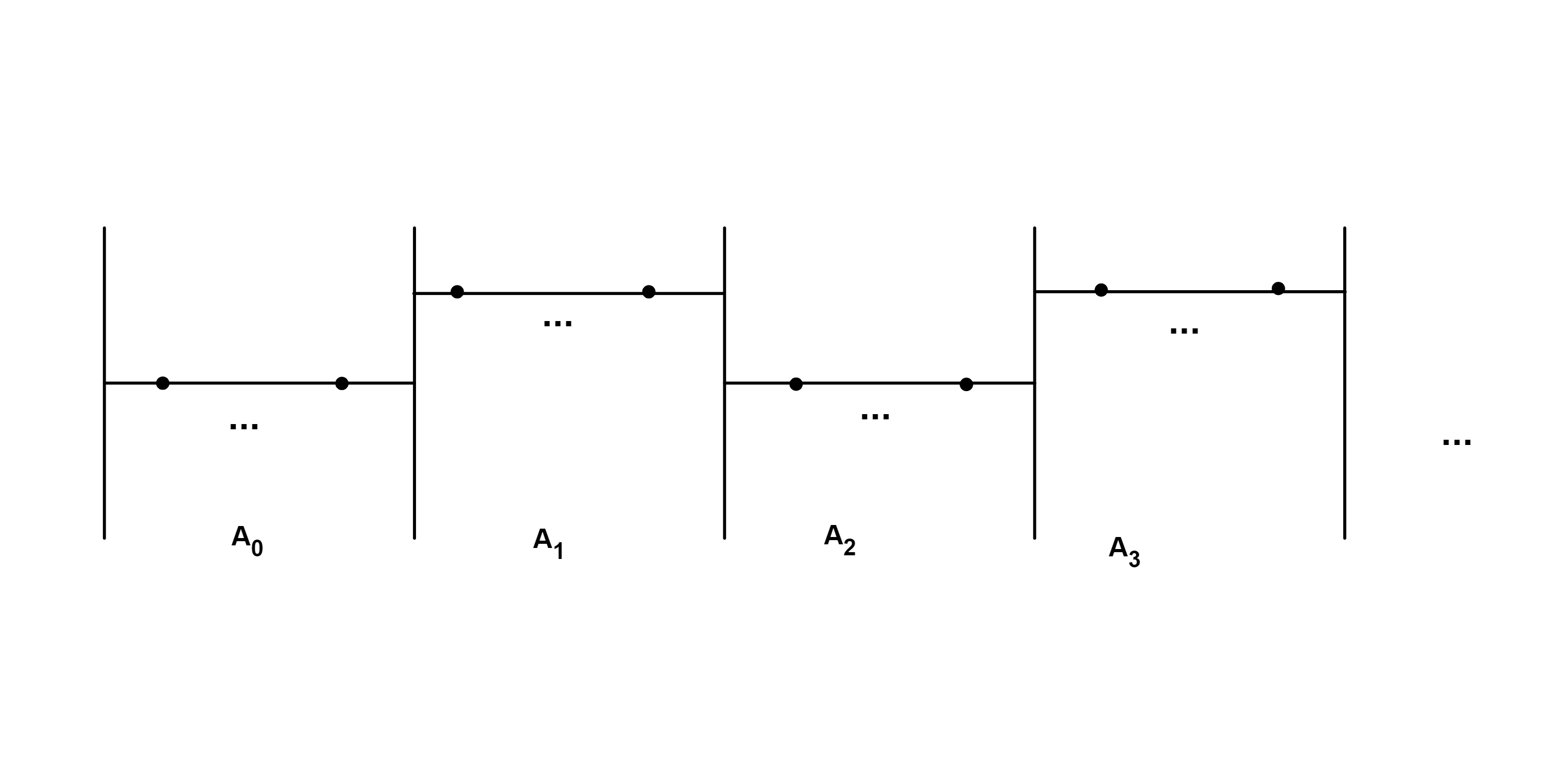}
}
\vspace{-1.3cm}
\caption{$S \cup T$ is partitioned into maximal subsets $A_0,A_1,A_2,\dots $.}
\label{fig:1}       
\end{figure}

In this section, we proceed with some useful definitions and assumptions. Let $S=\{s_1,s_2,\dots,s_y\}$ and $T=\{t_1,t_2,\dots,t_z\}$ be two point sets with $|S|+ |T|=n$. Let $D_S=\{\alpha_1,\alpha_2,\dots,\alpha_y\}$ and $D_T=\{\beta_1,\beta_2,\dots,\beta_z\}$ be the demand sets of $S$ and $T$, respectively. We denote the points in $S$ in increasing order by
$(s_1, . . . , s_y )$, and the points in $T$ in increasing order by $(t_1, . . . , t_z )$. Let $S \cup T$ be partitioned into maximal subsets $A_0,A_1,A_2,\dots $ alternating between subsets in $S$ and $T$ such that the largest point of $A_i$ lies to the left of the smallest point of $A_{i+1}$ for all $i\geq 0$ (see Figure \ref{fig:1}). W.l.o.g. we assume that all points $p \in S\cup T$ are distinct.

\begin{figure}[h]


\vspace{1cm}
\centering
\resizebox{1.3\width}{!}{%

  \includegraphics{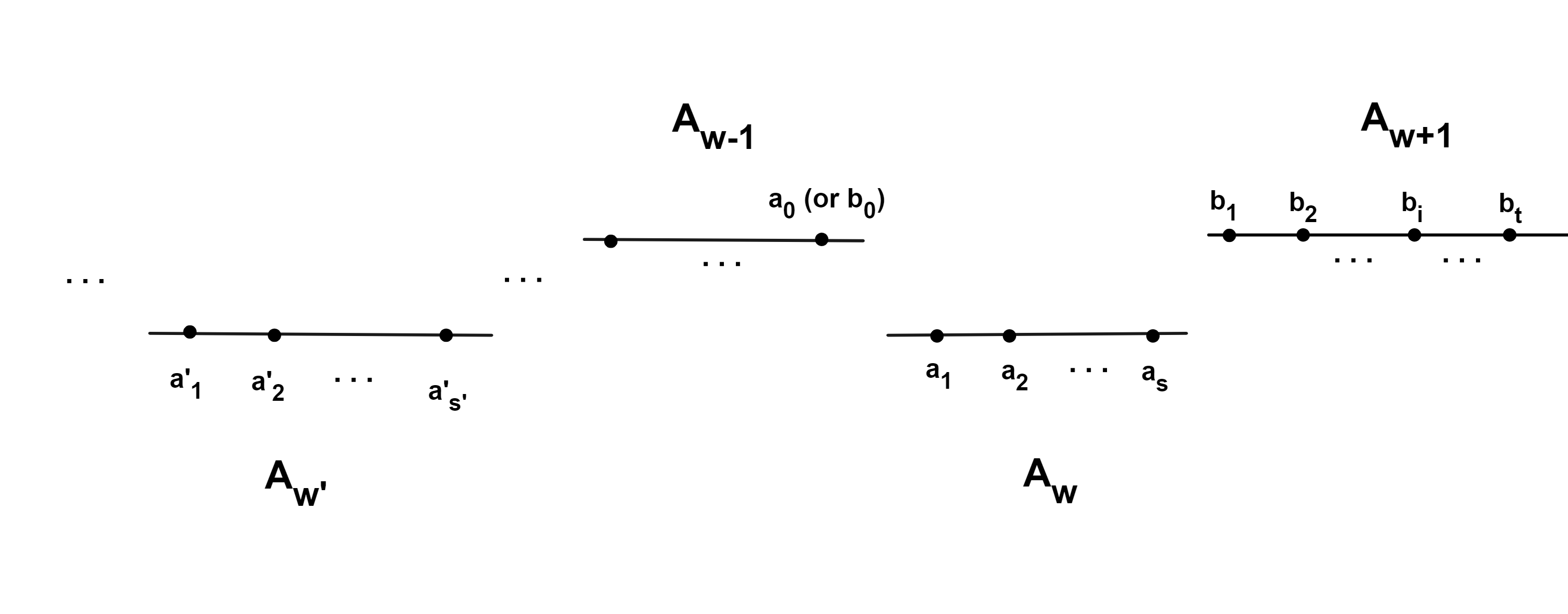}
}
\vspace{-0.5cm}
\caption{A minimum-cost OMMD between the points $p\leq a_s$ has been computed, and now we want to compute $C(b_i,k)$ for all $1 \leq i\leq t$ and $1 \leq k\leq k'$ for $Demand(b_i)\leq k'$.}
\label{fig:2}       
\end{figure}

Let $A_w=\{a_1,a_2,\dots,a_s\}$ with $a_1< a_2<\dots<a_s$ and $A_{w+1}=\{b_1,b_2,\dots,b_t\}$ with $b_1< b_2<\dots<b_t$. For $w>0$, $b_0$ or $a_0$ represents the largest point of $A_{w-1}$ (see Figure \ref{fig:2}). We denote the demand of each point $a \in S \cup T$ by $Demand(a)$. The cost of matching each point $a\in S$ to a point $b \in T$ is considered as $\Vert a-b\Vert$. For any point $q$, let $C(q,j)$ be the cost of a minimum-cost OMMD between the points $\{p \in S \cup T|p \leq q\}$, which satisfies all (or more) demands of each point $p$ with $p<q$ and exactly $j$ demands of $q$ (provided there exist enough points $p'\leq q$ for satisfying the demands). So, $C(q,0)$ is the cost of a minimum-cost OMMD between the points $\{p \in S \cup T|p \leq q\}$ such that $deg(p)\geq Demand(p)$ for all $p<q$ and $deg(q)=0$. And consequently, $C(b_i,0)=C(b_{i-1},deg(b_{i-1}))$ for $i>1$ and $C(b_1,0)=C(a_s,deg(a_s))$. Note that $deg(p)$ denotes the number of points that have been matched to $p$ in the OMMD. Let $M(b_{i},k)$ denote the point that satisfies the $k$th demand of $b_i$. Note that if $k>Demand(b_i)$, we suppose that $M(b_{i},k)$ is the $k$th point matched to $b_i$.

\section{Our algorithms}
\label{OMAsection}
In this section, we first present an $O(n^2)$ time algorithm based on the algorithm of \cite{Rajabi-Alni2} for finding a minimum-cost OMMD between two sets $S$ and $T$ lying on the real line. Then, we generalize our algorithm for computing an OMMDC between $S$ and $T$.

\subsection{The OMMD algorithm}
We first begin with some useful lemmas.


\begin{lemma}
\label{lem6}
Let $b<c$ be two points in $S$, and $a<d$ be two points in $T$ such that $a\le b<c\le d$. If a minimum-cost OMMD, denoted by $M$, contains both of $(a,c)$ and $(b,d)$, then $(a,b) \in M$ or $(c,d) \in M$.
\end{lemma}



\begin{figure}[h]

\vspace{1cm}
\hspace{1cm}
\resizebox{0.65\width}{!}{%

  \includegraphics{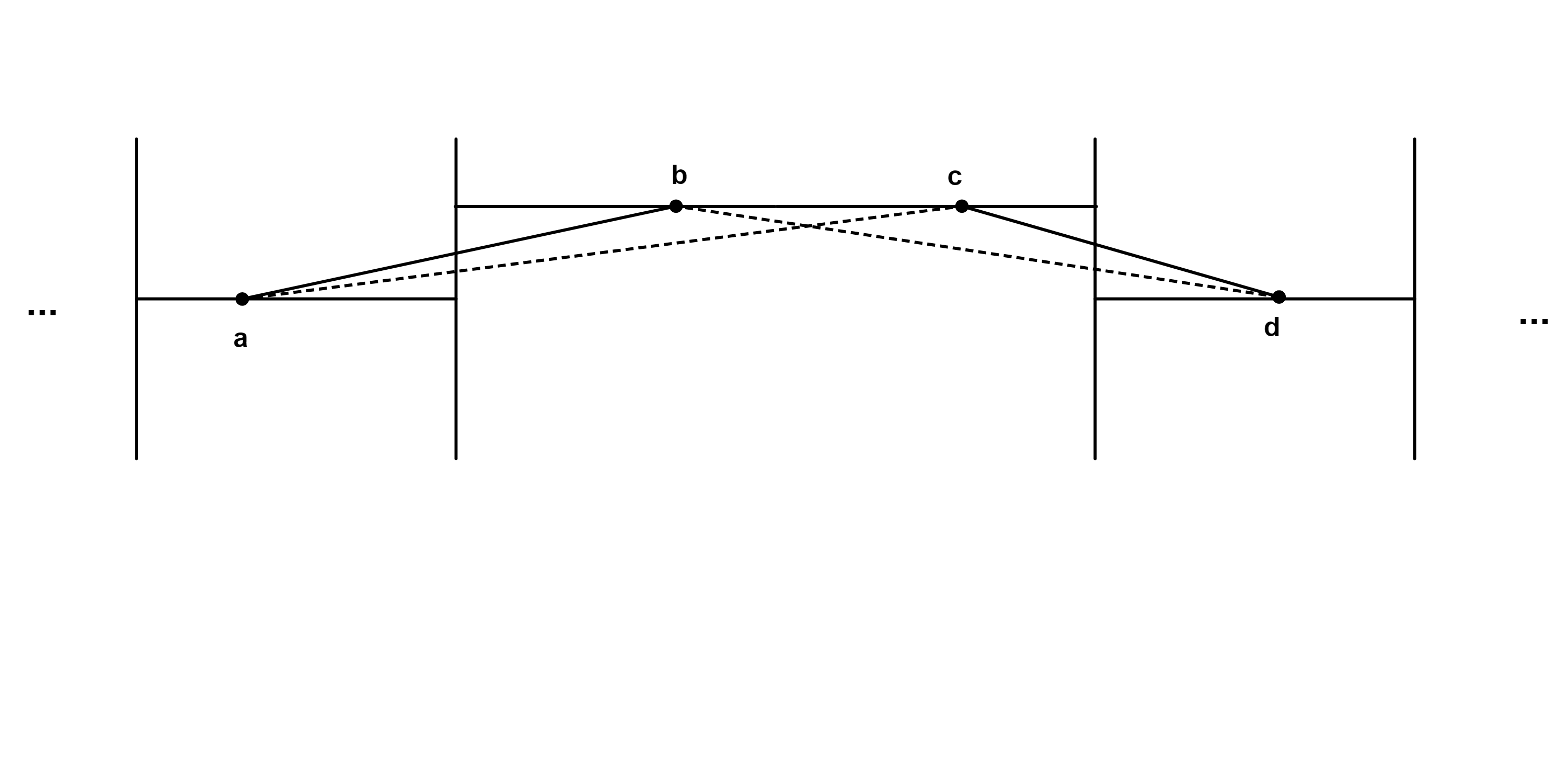}
}
\vspace{-2cm}
\caption{$(a,c)$ and $(b,d)$ do not both belong to an optimal matching.}
\label{fig:3}       
\end{figure}

\noindent \textbf{Proof.} Suppose that the lemma is false, and $M$ contains both $(a,c)$ and $(b,d)$, but neither $(a,b) \in M$ nor $(c,d) \in M$ (see Figure \ref{fig:3}). Then, we can remove the pairs $(a,c)$ and $(b,d)$ from $M$ and add the pairs $(a,b)$ and $(c,d)$: the result $M'$ is still an OMMD which has a smaller cost, a contradiction.\qed

\begin{lemma}
\label{lem5}
Let $a\in S,b\in T$ and $c\in S,d\in T$ such that $a\leq b<c\leq d$. Let $M$ be a minimum-cost OMMD. If $(a,d) \in M$, then either $(a,b) \in M$ or $(c,d) \in M$ or both.
\end{lemma}

\begin{figure}[h]

\vspace{1cm}
\hspace{1cm}
\resizebox{0.6\width}{!}{%

  \includegraphics{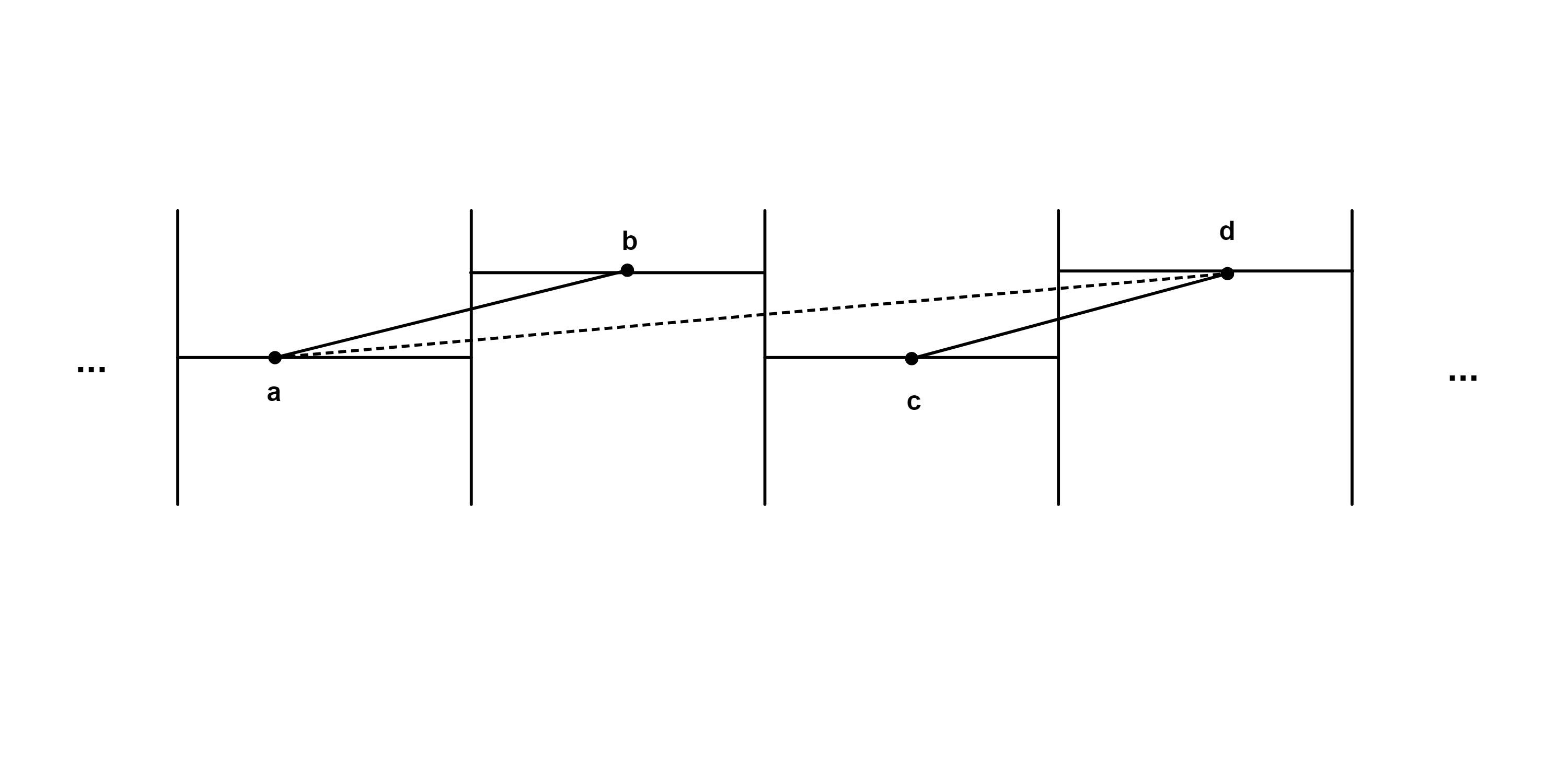}
}
\vspace{-2cm}
\caption{If $(a,d) \in M$, then either $(a,b) \in M$ or $(c,d) \in M$ or both.}
\label{fig:4}       
\end{figure}

\noindent \textbf{Proof.} Suppose for a contradiction that $(a,d) \in M$, but neither $(a,b) \in M$ nor $(c,d) \in M$. Then, if we remove the pair $(a, d)$ from $M$ and add the pairs $(a, b)$ and $(c, d)$, we get an OMMD with a smaller cost (see Figure \ref{fig:4}). Contradiction .\qed

\begin{figure}[h]

\vspace{1cm}
\hspace{1cm}
\resizebox{0.65\width}{!}{%

  \includegraphics{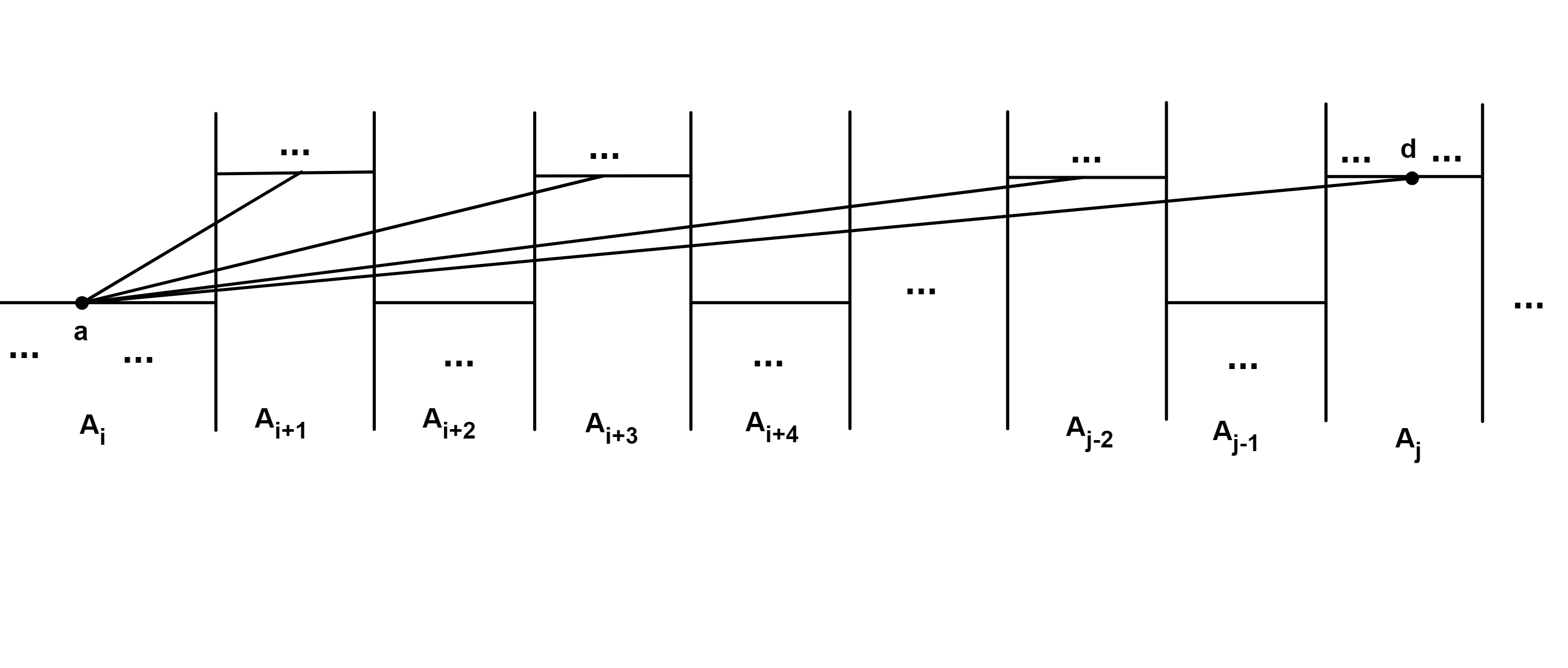}
}
\vspace{-1.5cm}
\caption{Illustration of the situation $(a,b) \in M$ for all $b \in A_{i+1} \cup A_{i+3}\cup  \dots \cup A_{j-2}$. The sets matched in a minimum-cost OMMD are connected with a line.}
\label{fig:5}       
\end{figure}

\begin{figure}[h]

\vspace{1cm}
\hspace{1cm}
\resizebox{0.65\width}{!}{%

  \includegraphics{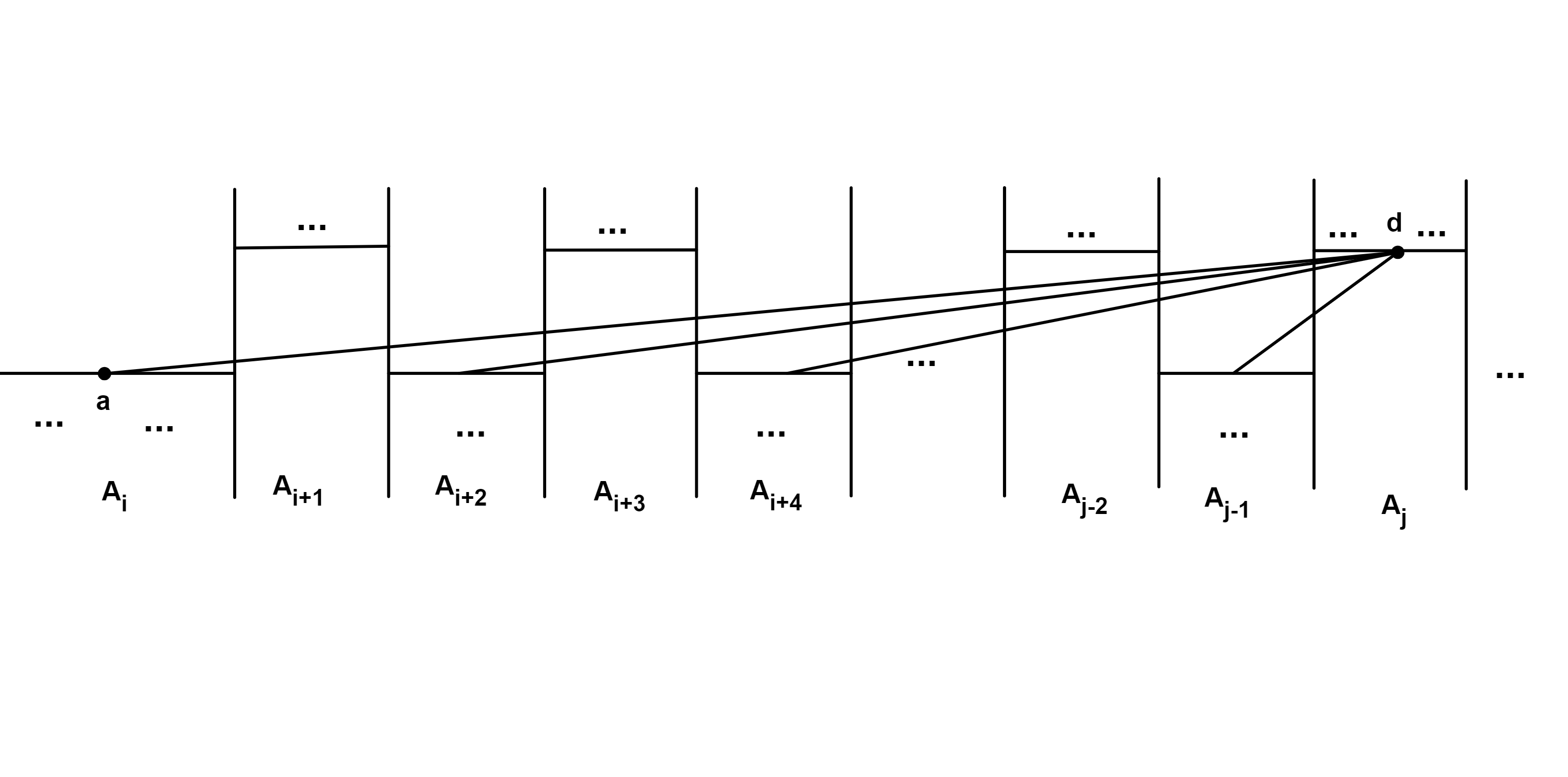}
}
\vspace{-2cm}
\caption{Illustration of the situation $(c,d) \in M$ for all $c \in A_{i+2} \cup A_{i+4}\cup  \dots \cup A_{j-1}$. The sets matched in a minimum-cost OMMD are connected with a line.}
\label{fig:6}       
\end{figure}

\begin{corollary}
\label{corlem5}
Let $a \in A_i$ and $d \in A_j$ for some $i,j \geq 0$. For any pairing $(a,d)$ in a minimum-cost OMMD, if $j>i+1$, then either $(a,b) \in M$ for all points $b \in A_{i+1} \cup A_{i+3}\cup  \dots \cup A_{j-2}$ (Figure \ref{fig:5}), or $(c,d) \in M$ for all points $c \in A_{i+2} \cup A_{i+4}\cup  \dots \cup A_{j-1}$ (Figure \ref{fig:6}) \citep{Rajabi-Alni}.
\end{corollary}

Note that we use Corollary \ref{corlem5} for satisfying the demands of $a \in A_i$ by the points of the sets $A_{i+1},A_{i+3},\dots, A_j$ or for satisfying the demands of $d \in A_j$ by the points of the sets $A_{j-1},A_{j-3}, \dots, A_i$.

\begin{theorem}
Let $S$ and $T$ be two sets of points on the real line with $\left|S\right|+\left|T\right|=n$. Then, a minimum-cost OMMD between $S$ and $T$ can be determined in $O(n^2)$ time.
\end{theorem}

\noindent \textbf{Proof.} Our algorithm is as follows (see Algorithm \ref{OMMD}). Initially, partition $S \cup T$ into maximal subsets $A_0,A_1,\dots$, and let $C(p,k)=\infty$ for all $p \in S \cup T$ and $1\leq k\leq n$ (Lines 1--5). Obviously, the number of the pairs of a minimum-cost OMMD is equal to $\max(\sum_{i=1}^{|S|}\alpha_i,\sum_{j=1}^{|T|}\beta_j)$. Note that we assume there exists at least one OMMD between $S$ and $T$.

Let $A_w=\{a_1,a_2, \dots, a_s\}$ and $A_{w+1}=\{b_1,b_2, \dots, b_t\}$ (Line 8). Assume that we have computed $C(p,h)$ for all $p < b_i$ and $1 \leq h\leq n$ (including $C(a_s,deg(a_s))$, i.e. the cost of an OMMD for the points $\{p\in S\cup T|p\leq a_s\}$). And now, we want to compute $C(b_i,k)$ for all $1 \leq i\leq t$ and $1\leq k\leq k'$, satisfying $Demand(b_i) \leq k'$ (see Figure \ref{fig:2}). In fact, we use the idea of \cite{Rajabi-Alni2}, that is we first examine the point $b_1 \in A_{w+1}$ and determine whether matching the points $p\leq b_1$ to the point $b_1$ decreases the cost of the OMMD or not. Then, if $deg(b_1) < Demand(b_1)$, we should satisfy the remaining demands of $b_1$ such that the cost of the OMMD is minimized. In other words, we compute $C(b_1,k)$ for $k={1,2, \dots ,k'}$ for $Demand(b_1)\leq k'$, respectively. Then, we examine the point $b_2$ and compute $C(b_2,k)$ for $k={1,2, \dots ,k'}$ for $Demand(b_2)\leq k'$, respectively, and so on. So, our algorithm is in two steps (Lines 7--23 of Algorithm \ref{OMMD}):

\begin{itemize}
\item [Step 1.]In this step, by Corollary \ref{corlem5}, we should determine whether matching the points $ p\leq b_i$ with $deg(p)\leq Demand(p)$ to $b_i$ decreases the cost of the OMMD or not (Line 11 of Algorithm \ref{OMMD}). We distinguish two cases:
    \begin{itemize}

      \item $deg(p)= Demand(p)$ and $p$ has been matched to at least a point $a'\leq p$. Let $q$ denote the number of points $a'\leq p$ that have been matched to $p$. If we have $C(p,q)>C(p,q-1)+b_i-p$, we remove the pair $(p,M(p,q))$ from the OMMD and add the pair $(p,b_i)$ to it.
      \item $deg(p)< Demand(p)$. Then, we add the pair $(p,b_i)$ to the OMMD.
    \end{itemize}

    Let $list(w)$ denote a doubly linked list of the points $p\in A_w$, which is initially empty (Line 9 of Algorithm \ref{OMMD}). We construct $list(w)$ such that $Match(b_i,w)$ denotes a pointer to the sublist of points in $list(w)$ that are matched to $b_i$. More precisely, $Match(b_i,w)$ points to the end of a doubly linked list of the points of $A_w$ matched to $b_i$ (starting from $list(w)$.first when $Match(b_i,w)\neq null$). Initially, $Match(b_i,w)=null$ (Line 1 of Algorithm \ref{Step1}).
    
    Firstly, for $i=1$, we start from $a_s$ and examine the points of $A_w$, i.e., $a_s,a_{s-1}, \dots, a_1$, respectively to determine whether matching each point $a_j \in A_w$ to $b_1$ decreases the cost of the OMMD or not until reaching the point $a_0$ (Lines 2--16 of Algorithm \ref{Step1}; see Figure \ref{fig:2}). Observe that when a point $a_j$ is matched to $b_1$, we add $a_j$ to the end of $list(w)$ (Lines 9 and 14 of Algorithm \ref{Step1}). Finally, we update $Match(b_1,w)$ (Line 16 of Algorithm \ref{Step1}). Note that $list(w).first$ and $list(w).end$ are pointers to the first and last elements of $list(w)$, respectively.

    Note that if matching a point $p\leq b_{i-1}$ to $b_{i-1}$ does not decrease the cost of the OMMD, then matching $p$ to $b_i$ does not either. Thus, for $i\geq 1$ we do as follows. Let $tempset$ denote the set of the indices of the partitions $A_{w'}$ containing at least one point matched to $b_{i-1}$, that is $tempset=\{w':M(b_{i-1},k)\in A_{w'}\}_{k=1}^{deg(b_{i-1})}$, in descending order (Line 1 of Algorithm \ref{Step1:2}). We must verify whether matching the points $M(b_{i-1},k)$ to $b_i$ decreases the cost of the OMMD or not. Thus, for each $w' \in tempset$ (starting from the largest index $w''$), we search the doubly linked list whose first and last pointers are $list(w').first$ and $Match(b_{i-1},w')$, respectively; this list is denoted by $list(w',first:Match(b_{i-1},w'))$ (Lines 20--36 of Algorithm \ref{Step1:2}). 
    
    We construct two linked lists, $templist1$ and $templist2$, of the points $a'_j\in A_{w'}$. $templist1$ contains points whose matching to $b_i$ decreases the cost of the OMMD (Lines 27 and 33 of Algorithm \ref{Step1:2}), while $templist2$ holds points whose matching to $b_i$ does not decrease the cost (Lines 29 and 35 of Algorithm \ref{Step1:2}). Then, we concatenate $templist1$ and $templist2$ using the function $Concatenate(templist1,templist2)$ to get a doubly linked list $templist3$ where $templist1.end.next=templist2.first$ and $templist2.first.prev=templist1.end$ (Line 37 of Algorithm \ref{Step1:2}). And, we also copy $templist3$ into $list(w'',first:Match(b_{i-1},w''))$ (Line 39 of Algorithm \ref{Step1:2}). Assume that $k'=num(templist1)$ denotes the number of the points of $templist1$ (Line 38 of Algorithm \ref{Step1:2}). Finally, we set $Match(b_i,w'')$ to point to the $k'$th point of $list(w'')$, which is denoted by $list(w'',k')$ (Line 40 of Algorithm \ref{Step1:2}).

    Note that this step can be considered a preprocessing step for each point $b_i \in A_{w+1}$, and has the time complexity of $O(n)$. Since the number of the points $p\leq b_i$ in $S\cup T$ is at most $O(n)$. Then, if $deg(b_i)\geq Demand(b_i)$, we are done. Otherwise, we go to Step 2. Note that if $deg(b_i)> Demand(b_i)$, we add $b_i$ to the linked list $LL(w+1)$ (Lines 12--13 of Algorithm \ref{OMMD}).


\item [Step 2.]This step consists of two substeps, Step 2.1 and Step 2.2, which are performed iteratively until $deg(b_i)>Demand(b_i)$ or $w'\leq 0$ (Lines 16--22 of Algorithm \ref{OMMD}). Note that initially $w'=w+1$ (Line 15 of Algorithm \ref{OMMD}).

      \begin{itemize}
        \item [Step 2.1]In this step, we should check the points $p\in A_{w'}$ with $p<b_i$ in descending order to find the first (i.e. the largest) point $b'_j\in A_{w'}$ with $deg(b'_j)> Demand(b'_j)$. Thus, we use the list $LL(w')$ which maintains the points $p\in A_{w'}$ satisfying $deg(p) > Demand(p)$ (Lines 12--13 of Algorithm \ref{OMMD}). Note that $LL$ is a set of lists, one list for each partition $A_{j}$ for $j\geq 1$.

            Let $templist=LL(w')$ and $tempset=\emptyset$ (Line 1 of Algorithm \ref{Step2}). While $templist\neq \emptyset$ and $deg(b_i)<Demand(b_i)$, we search $A_{w'}$ as follows (Lines 2--18 of Algorithm \ref{Step2}). Firstly, if $tempset=\emptyset$, we remove the last (the largest) point of $templist$ denoted by $b'_j$ (Line 4 of Algorithm \ref{Step2}). Let $tempset=\{w'':M(b'_j,k)\in A_{w''}\}_{k=1}^{deg(b'_j)}$ such that the elements of $tempset$ are in ascending order (Line 5 of Algorithm \ref{Step2}).

      Remove the smallest index $w''$ from $tempset$ (Line 6 of Algorithm \ref{Step2}). Let $current=Match(b'_j,w'')$ (Line 7 of Algorithm \ref{Step2}). While
      \begin{itemize}
        \item $Match(b_i,w'')\neq current$,
        \item and $deg(b_i)<Demand(b_i)$,
        \item and $deg(b'_j)>Demand(b'_j)$,
      \end{itemize}
         starting form $current$, we seek $list(w'')$ as follows (Lines 8--17 of Algorithm \ref{Step2}). Let $p'$ be the value of the element whose pointer is $current$ in $list(w'')$ (Line 9 of Algorithm \ref{Step2}). We should match $b_i$ to $p'$ and remove $(b'_j,p')$ from the OMMD (Lines 10--12 of Algorithm \ref{Step2}). Since otherwise, $b_i$ would be matched to a point $p$ for which one of the following statements holds:
    \begin{itemize}
      \item either $p\in \{M(b'_j,v)\}_{v=1}^{deg(b'_j)}$ (see Figure \ref{fig:7}). Then, if we remove the pair $(p,b'_j)$ from the OMMD, we get an OMMD with a smaller cost.
      \item or $p\notin \{M(b'_j,v)\}_{v=1}^{deg(b'_j)}$ (see Figure \ref{fig:8}). Then, two cases arise:

      \begin{itemize}
        \item either $deg(p)=demand(p)$. Note that $deg(b'_j)> Demand(b'_j)$ implies that matching the point $M(b'_j,v)$ to $b'_j$ for $v\in \{1,2,\dots,deg(b'_j)\}$ decreases the cost of the OMMD, and matching $p$ to $b'_j$ does not decrease the cost of the OMMD. Therefore, we have:
      $$C(p,v')<b'_j-p+C(p,v'-1),$$ and
      $$0<b'_j-p+C(p,v'-1),$$
      for all $1\leq v'\leq deg(p)$.
      Therefore,
      $$b_i-b'_j<b_i-b'_j+b'_j-p+C(p,v'-1).$$ Thus, we have
      $$C(b_i,k-1)+b_i-b'_j<C(b_i,k-1)+b_i-p+C(p,v'-1),$$ which means if we add $(M(b'_j,v),b_i)$ to the OMMD and remove $(M(b'_j,v),b'_j)$ for an arbitrary $v\in \{1,2,\dots,deg(b'_j)\}$, say $v''$, we get a smaller cost. We assume w.l.o.g that $M(b'_j,v'')=\min^{deg(b'_j)}_{v=1}(M(b_j,v))$.
        \item or $deg(p)>demand(p)$. This case implies that
        $$b_i-p<b_i-b'_j,$$ contradicting $p<b'_j$.
      \end{itemize}

      \end{itemize}

   Then, we update $current$ (Line 13 of Algorithm \ref{Step2}). Observe that if $deg(b'_j)=Demand(b'_j)$, we remove $b'_j$ from $LL(w')$ and let $tempset=\emptyset$ (Lines 15--17 of Algorithm \ref{Step2}).


   If there does not exist such a point $b'_j\leq b_i$ with $deg(b'_j)> Demand(b'_j)$, by the following claim, we check whether the points $b'_j \in A_{w'}$ have been matched to any points $p$ with $deg(p)> Demand(p)$, which are maintained in the linked list $LL2(w')$ (Lines 19--25 of Algorithm \ref{Step2}). Note that $LL2(w')$ is a linked list of the points of $\{M(b'_{j},v):deg(M(b'_{j},v))> Demand(M(b'_{j},v))\}^{Demand(b'_{j})}_{v=1}$ for $1\leq j\leq t$, in decreasing order (Lines 12--13 of Algorithm \ref{Step3}). Therefor:
$$a'_h=\max(\{M(b'_{j},v):deg(M(b'_{j},v))> Demand(M(b'_{j},v))\}^{Demand(b'_{j})}_{v=1}).$$
Then, if $a'_h$ has not been matched to $b_i$, we match $b_i$ to $a'_h$ (Lines 23--25 of Algorithm \ref{Step2}).
It is easy to show that thank the linked lists $LL(w')$ and $LL2(w')$, Step 2.1 runs in $O(n)$ time for each $b_i\in A_{w+1}$.

\newtheorem{claim}{Claim}
\begin{claim}
\label{cccc}
Let $a'_h$ be the largest point with $deg(a'_h)> Demand(a'_h)$ that has been matched to $b'_j$ in $A_{w'}$, then $b_i$ is also matched to $a'_h$.
\end{claim}

\noindent \textbf{Proof.} Note that $deg(a'_h)> Demand(a'_h)$ implies that matching some of the points to $a'_h$ decreases the cost of the OMMD. Also note that $\Vert b'_j-a\Vert<\Vert b_i-a\Vert$ for all points $a \in S \cup T$ with $a\leq b'_j$. In the minimum-cost OMMD, $b'_j$ has been matched to $a'_h=M(b'_j,v)$ for $v\in \{1,2,\dots, Demand(b'_j)\}$ instead of any other point $a'_q$ with $deg(a'_q)\geq Demand(a'_q)$ (see Figure \ref{fig:5}), so:

\begin{eqnarray}
  \nonumber C(b'_{j},v-1)+b'_j-a'_h<C(b'_{j},v-1)+b'_j-a'_q  & \\
 \nonumber + \min(-C(a'_q,deg(a'_q))+C(a'_q,deg(a'_q)-1),0),&
\end{eqnarray}

and thus:

$$-a'_h<-a'_q+ \min(-C(a'_q,deg(a'_q))+C(a'_q,deg(a'_q)-1),0),$$

If we add $C(b_i,k-1)$ and $b_i$ to both sides of the above inequality, then we have:

\begin{eqnarray}
  \nonumber C(b_i,k-1)+b_i-a'_h<C(b_i,k-1)+b_i-a'_q  & \\
 \nonumber + \min(-C(a'_q,deg(a'_q))+C(a'_q,deg(a'_q)-1),0),&
 \end{eqnarray}

so $b_i$ is also matched to $a'_h$.\qed

\begin{figure}[h]

\vspace{1cm}
\hspace{1cm}
\resizebox{1.5\width}{!}{%

  \includegraphics{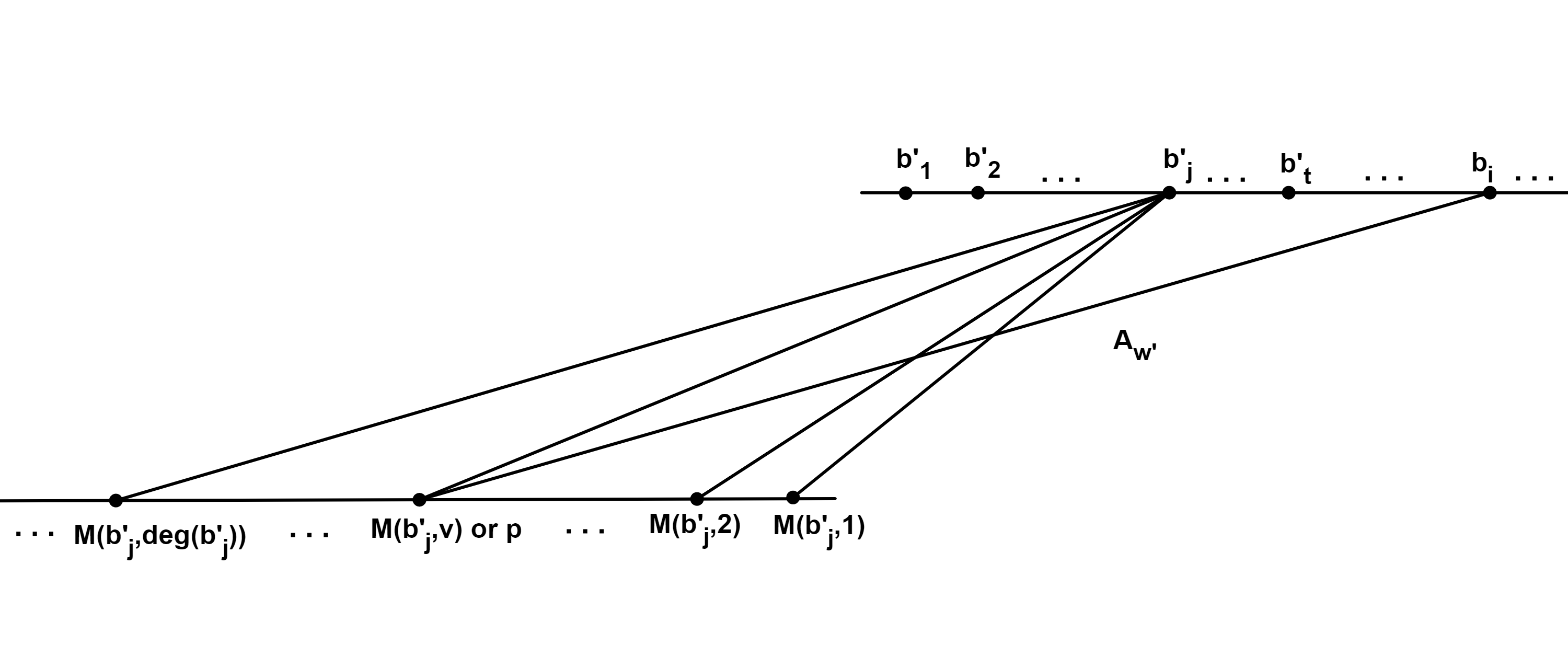}
}
\vspace{-0.5cm}
\caption{Illustration of the case where $b_i$ has been matched to a point $p\in \{M(b_j,v)\}_{v=1}^{deg(b_j)}$.}
\label{fig:7}       
\end{figure}

\begin{figure}[h]

\vspace{1cm}
\hspace{1cm}
\resizebox{1.6\width}{!}{%

  \includegraphics{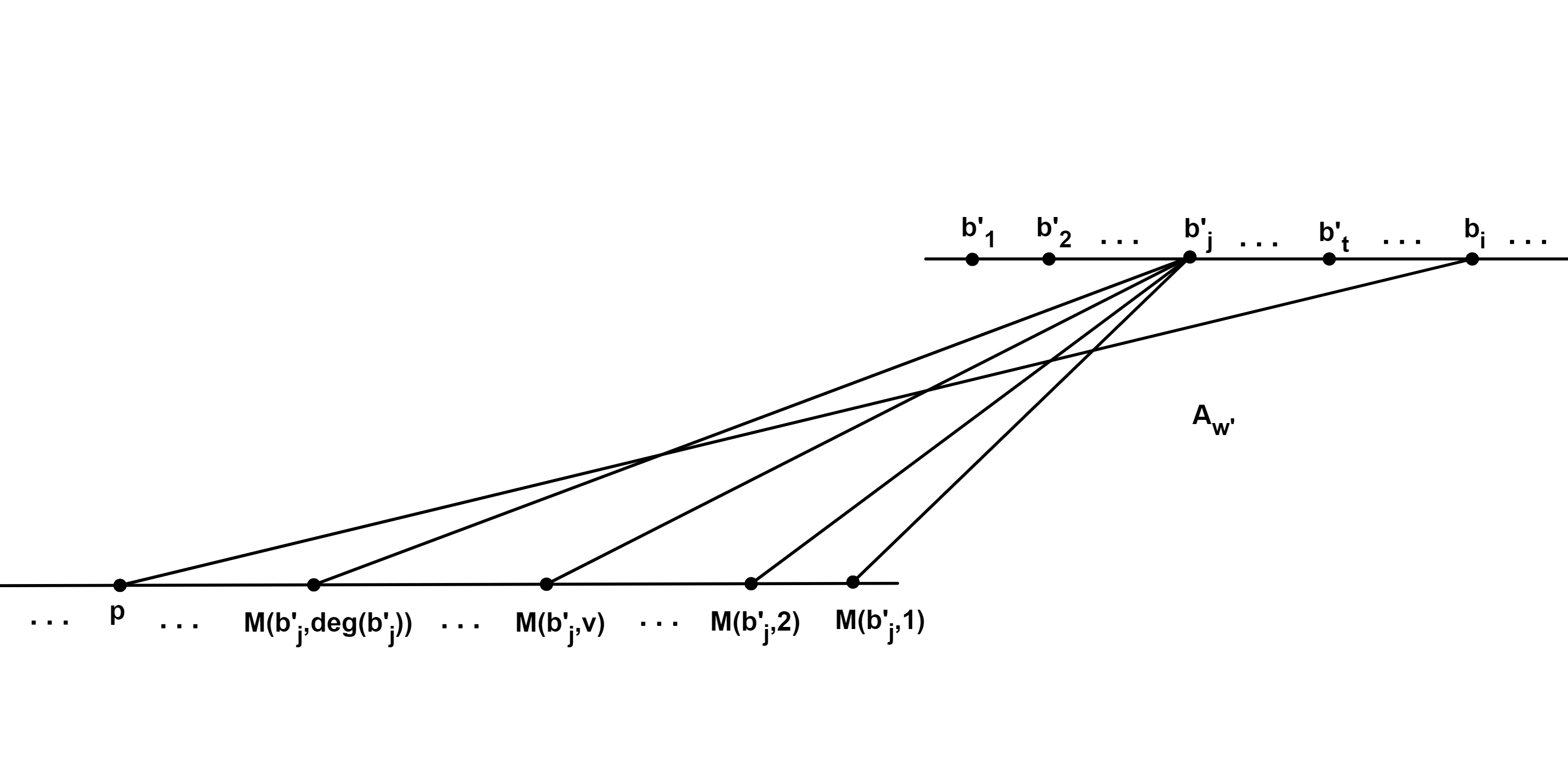}
}
\vspace{-0.5cm}
\caption{Illustration of the case where $b_i$ has been matched to a point $p\notin \{M(b_j,v)\}_{v=1}^{deg(b_j)}$.}
\label{fig:8}       
\end{figure}

If $deg(b_i)\geq Demand(b_i)$ we have done. Otherwise, we let $w'=w'-1$ and go to Step 2.2 (Lines 18--20 of Algorithm \ref{OMMD}).


\item [Step 2.2] In this step, by Corollary \ref{corlem5}, we seek the partition $A_w'$ to satisfy the demands of $b_i$ (Algorithm \ref{Step3}; see Figure \ref{fig:2}). Let $A_{w'u}\neq \emptyset$ be the set of the points in $A_{w'}$ that have been matched to $u\geq 0$ smaller points. Let $R$ denote the set of the numbers $u\geq 0$ with $A_{w'u}\neq \emptyset$ and $a''_u$ be the largest point of $A_{w'u}$.
\begin{claim}
\label{lemma_main1}
Assume that $a'_l,a'_{m}\in A_{w'u}$ with $a'_{m}< a'_l$. In this step, $b_i$ can be matched to $a'_l$ but not to $a'_{m}$.
\end{claim}


\noindent \textbf{Proof.} Suppose by contradiction that the point $b_i$ is matched to $a'_{m}$ (the pair $(M(a'_m, u),a'_m)$ might be removed from the OMMD and the pair $(a'_m,b_i)$ is added to the OMMD). Thus, two cases arise:
\begin{itemize}
  \item $(M(a'_m, u),a'_m)$ is not removed. Then, if we remove $(a'_m,b_i)$ from the OMMD and add $(a'_l,b_i)$, we get an OMMD with a smaller cost. Contradiction.
  \item $(M(a'_m, u),a'_m)$ is removed. Then, it is easy to show that there exist at least one point $b'\leq a'_m$ such that $(b',a'_l)\in OMMD$ but $(b',a'_m)\notin OMMD$; by Lemma \ref{lem6}, two pairs $(b',a'_l)$ and $(a'_m,b_i)$ contradict the optimality (see Figure \ref{fig:9}). \qed
\end{itemize}

Thus, by Claim \ref{lemma_main1}, we determine that $b_i$ should be matched to which of the points $a''_u$ for $u \in R$. Note that if we match $a''_u$ to $b_i$, we must check that whether the OMMD still includes the pair $(M(a''_u, u),a''_u)$ or not, i.e. it is possible that after matching $a''_u$ to $b_i$, the pair $(M(a''_u, u),a''_u)$ is removed from the OMMD.

Let $u(p)$ be the number of the points $M(p,h)$ with $M(p,h)<p$, i.e., $$u(p)=|\{M(p,h):M(p,h)<p\}_{h=1}^{deg(p)}|,$$ for $p\in A_{w'}$ (Lines 1--2 of Algorithm \ref{Step3}). Assume $templist$ denotes the list of the points $p\in A_{w'}$ in ascending order of
$$C(b_i,k-1)+b_i-p+\min(-C(p,u(p))+C(p,u(p)-1),0),$$
i.e., the cost of matching $b_i$ to $p$ (Line 4 of Algorithm \ref{Step3}). Let $templist=templist-\{p\in A_{w'}:(p,b_i)\in OMMD\}$ (Line 5 of Algorithm \ref{Step3}). While $templist\neq \emptyset$ and $deg(b_i)<Demand(b_i)$, we remove the first point of $templist$ denoted by $p'$ (the point that matching $b_i$ to it has the smallest cost), and match $b_i$ to $p'$ (Lines 8--9 of Algorithm \ref{Step3}). Then, we remove the pair $(M(p',u(p')),p')$ from the OMMD, if it is necessary (Lines 10--11 of Algorithm \ref{Step3}). Now, if $deg(p')> Demand(p')$, we add $p'$ to the end of the linked list $LL2({w'})$ which later might be used in Step 2.1 (Lines 15--16 of Algorithm \ref{Step3}).

\end{itemize}

This step runs in $O(n)$ time, since the number of the points $p\leq b_i$ is at most $n$.\qed

\end{itemize}
\begin{figure}[h]

\vspace{1cm}
\hspace{2.7cm}
\resizebox{1\width}{!}{%

  \includegraphics{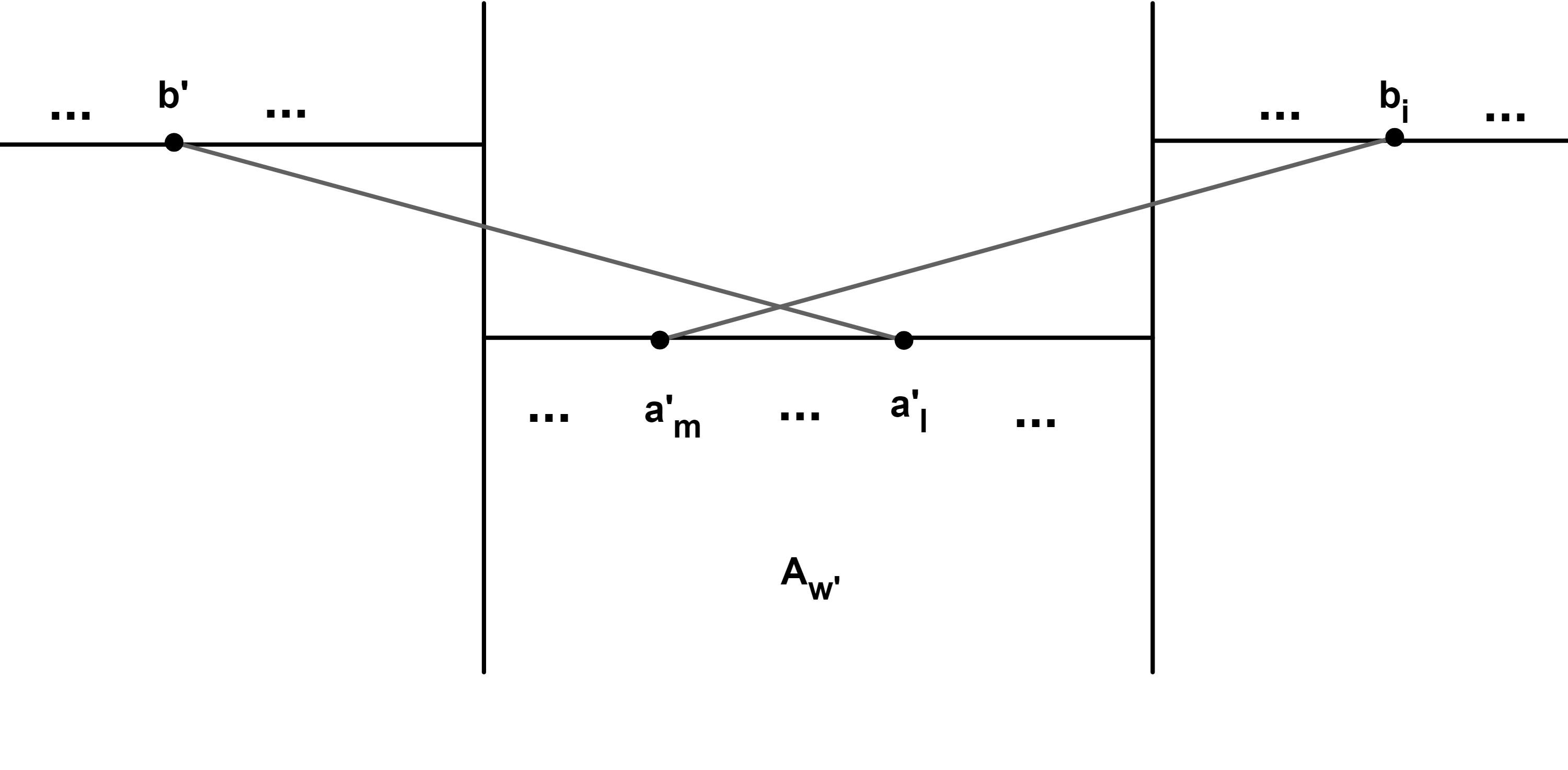}
}
\vspace{0cm}
\caption{Illustration of proof of Claim \ref{lemma_main1}.}
\label{fig:9}       
\end{figure}


\vspace{3cm}
\begin{algorithm}[]
\caption{OMMD Algorithm}
\label{OMMD}

\SetKwInOut{KwIn}{Input}
\SetKwInOut{KwOut}{Output}

\KwIn{$S$, $T$, $D_S$, $D_T$}
\KwOut{Processed list.}
Partition $S \cup T$ to $A_0,A_1, \dots $\;
Let $n=|S \cup T|$\;
\ForAll {$p \in S \cup T$}{
\ForAll {$1\leq k \leq n $}{
$C(p,k)=\infty$\;
}
}

Let $w=0$ and $OMMD=\emptyset$\;
\While{$|OMMD|<\max(\sum_{i=1}^{|S|}\alpha_i,\sum_{j=1}^{|T|}\beta_j)$}{

 Let $A_w=\{a_1,a_2, \dots, a_s\}$, and $A_{w+1}=\{b_1,b_2, \dots, b_t\}$\;
 Let $list(w)$ be an empty linked list\;
 \For {$i=1$ to $t$}{
 \textbf{Step1}($S,T,A_0,A_1,\dots$)\;

\If{$deg(b_i)> Demand(b_i)$}{
 Add $b_i$ to the end of the linked list $LL({w+1})$\;
}


}

\For {$i=1$ to $t$}{

 Let $w'=w+1$\;

\While{$deg(b_i)<Demand(b_i)$ and $w'\geq 1$}{

\textbf{Step 2.1}($S,T,A_0,A_1,\dots,w',b_i$)\;

\If{$deg(b_i)< Demand(b_i)$}{

Let $w'=w'-1$\;
\textbf{Step 2.2}($S,T,A_0,A_1,\dots,w',b_i$)\;

}

\If{$deg(b_i)< Demand(b_i)$}{

Let $w'=w'-1$\;

}

}

}

$w=w+1$\;
}

 \end{algorithm}

\begin{algorithm}[]

\caption{Step1:PartI($S$,$T$,$A_0$,$A_1$,$\dots$)}
\label{Step1}

Let $Match(b_i,w)=null$\;
\If {$i=1$}{
 $j=s$\;
\While{$j\geq 1$}{
 Let $q$ be the number of the points of $\{M(a_j,h):M(a_j,h)\leq a_j\}_{h=1}^{deg(a_j)}$\;

\If{$deg(a_j)= Demand(a_j)$ and $q\neq 0$}{

\If {$C(a_j,q)>C(a_j,q-1)+b_1-a_j$}{
 Add the pair $(b_1,a_j)$ to OMMD and remove $(a_j,M(a_j,q))$\;
 Add $a_j$ to the end of $list(w)$\;
 $C(b_1,deg(b_1))=C(b_1,deg(b_1)-1)+b_1-a_j-C(a_j,q)+C(a_j,q-1)$\;

}

}

\ElseIf{$deg(a_j)< Demand(a_j)$}{
 Add the pair $(b_1,a_j)$ to OMMD\;
 $C(b_1,deg(b_1))=C(b_1,deg(b_1)-1)+b_1-a_j$\;
 Add $a_j$ to the end of $list(w)$\;
}

 $j=j-1$\;
}
Let $Match(b_i,w)=list(w).end$\;

}

 \end{algorithm}

\begin{algorithm}[]

\caption{Step1:PartII($S$,$T$,$A_0$,$A_1$,$\dots$)}
\label{Step1:2}
\setcounter{AlgoLine}{15}
 Let $tempset$ denote the set $\{w':M(b_{i-1},k)\in A_{w'}\}_{k=1}^{deg(b_{i-1})}$ in descending order\;
 \While {$tempset\neq \emptyset$}{
 Remove the first (largest) index $w''$ from the set $tempset$\;
 Let $current=list(w'').first$\;
 \While {$current\neq Match(b_{i-1},w'').next$}{
 
 Let $a'_j=current.val$\;
 Let $q$ be the number of the points of $\{M(a'_j,h):M(a'_j,h)\leq a'_j\}_{h=1}^{deg(a'_j)}$\;
 \If{$deg(a'_j)= Demand(a'_j)$ and $q\neq 0$}{
 
 \If{$C(a'_j,q)>C(a'_j,q-1)+b_i-a'_j$}{
 Add the pair $(b_i,a'_j)$ to OMMD and remove $(a'_j,M(a'_j,q))$\;
 $C(b_i,deg(b_i))=C(b_i,deg(b_i)-1)+b_i-a'_j-C(a'_j,q)+C(a'_j,q-1)$\;
Add $a'_j$ to the end of $temlist1$\;

}

\Else{
Add $a'_j$ to the end of $templist2$\;
} 
 
 }
\ElseIf{$deg(a'_j)< Demand(a'_j)$}{
 Add the pair $(b_i,a'_j)$ to OMMD\;
 $C(b_i,deg(b_i))=C(b_i,deg(b_i)-1)+b_i-a'_j$\;
Add $a'_j$ to the end of $templist1$\;

}
 \Else{
Add $a'_j$ to the end of $templist2$\;
} 
 $current=current.next$\;
 }
 
 $templist3=Concatenate(templist1,templist2)$\;
 $k'=num(templist1)$\;
 Copy $templist3$ into $list(w'',first:Match(b_{i-1},w''))$\;
 $Match(b_i,w'')=list(w'',k')$\;

}
 \end{algorithm}

\begin{algorithm}[]

\caption{Step2.1($S$,$T$,$A_0$,$A_1$,$\dots$,$w'$,$b_i$)}
\label{Step2}

   Let $templist=LL(w')$ and $tempset=\emptyset$\;

   \While {$templist\neq \emptyset$ and $deg(b_i)<Demand(b_i)$}{

      \If{$tempset=\emptyset$}
      {
      Remove the last point of $templist$ denoted by $b'_j$\;
      Let $tempset$ denote the set $\{w'':M(b'_j,k)\in A_{w''}\}_{k=1}^{deg(b'_j)}$ in ascending order\;
      }

      Remove the first (smallest) index $w''$ from the set $tempset$\;
      Let $current=Match(b'_j,w'')$\;

      \While{$Match(b_i,w'')\neq current$ and $deg(b_i)<Demand(b_i)$ and $deg(b'_j)>Demand(b'_j)$}{
       Let $p'=current.val$ and assume $p'=M(b'_j,k')$\;
       \If{$k'\neq deg(b'_j)$}{
       Let $M(b'_j,k')=M(b'_j,deg(b'_j))$ and $M(b'_j,deg(b'_j))=p'$\;
       }
       Match $b_i$ to $p'$ and remove $(b'_j,p')$ from the OMMD\;
       Let $current=current.prev$\;
       Let $k=deg(b_i)$ and $C(b_i,k)=C(b_i,k-1)+b_i-b'_j$\;
      \If{$deg(b'_j)=Demand(b'_j)$}{
      Remove $b'_j$ from $LL(w')$\;
      Let $tempset=\emptyset$\;

      }
      }

     Let $Match(b_i,w)=Match(b'_j,w'')$ and $Match(b'_j,w'')=current$\;
     }

 Let $templist=LL2(w')$\;
\While{$templist\neq \emptyset$ and $deg(b_i)<Demand(b_i)$}{
 Let $k=deg(b_i)+1$\;
 Remove the first point of $templist$ called $a'_h$\;
\If{$(a'_h,b_i) \notin \ OMMD$}{
 Match $b_i$ to the point $a'_h$\;
 $C(b_i,k)=C(b_i,k-1)+b_i-a'_h$\;
}
}

 \end{algorithm}

\begin{algorithm}[]

\caption{Step2.2($S$,$T$,$A_0$,$A_1$,$\dots$,$w'$,$b_i$)}
\label{Step3}

\For{$p\in A_{w'}$}
{
Let $u(p)=|\{M(p,h):M(p,h)<p\}_{h=1}^{deg(p)}|$\;
}
Let $k=deg(b_i)+1$\;
Let $templist$ be the list of the points $p\in A_{w'}$ in ascending order of
$C(b_i,k-1)+b_i-p+\min(-C(p,u(p))+C(p,u(p)-1),0)$\;
Let $templist=templist-\{p\in A_{w'}:(p,b_i)\in OMMD\}$\;
\While {$deg(b_i)<Demand(b_i)$ and $templist \neq null$}{
 Let $k=deg(b_i)+1$\;
 Remove the first point of $templist$ denoted by $p'$\;
 Match $b_i$ to the point $p'$\;
 \If {$C(p',u(p'))>C(p',u(p')-1)$}{
 Remove $(M(p',u(p')),p')$ from $OMMD$\;
 $C(b_i,k)=C(b_i,k-1)+b_i-p-C(p,u(p))+C(p,u(p)-1)$\;
 }
\Else{
$C(b_i,k)=C(b_i,k-1)+b_i-p$\;
}

\If{$deg(p')> Demand(p')$}{

 Add $p'$ to the end of the linked list $LL2({w'})$\;
 
}


}

 \end{algorithm}

\subsection{The OMMDC algorithm}

Given two sets of points, $S$ and $T$, with $|S|+|T|=n$ on the real line, in this section, we provide an algorithm based on the OMMD algorithm for finding a minimum-cost OMMDC between $S$ and $T$ in $O(n^2)$ time. In an OMMDC, we should consider the limited capacities of points in addition to their demands. The correctness of the OMMDC algorithm follows from the proof of correctness for the OMMD algorithm, since the two algorithms are almost identical except for a slight modification to account for point capacities.
\begin{theorem}
Let $S$ and $T$ be two sets of points on the real line with $\left|S\right|+\left|T\right|=n$. Then, a minimum-cost OMMDC between $S$ and $T$ can be computed in $O(n^2)$ time.
\end{theorem}

\noindent \textbf{Proof.} We use the same notations as for Algorithm \ref{OMMD}. Suppose that we have computed $C(p,h)$ for all $\{p\in S\cup T|p<b_i\}$ and $1\leq h\leq Cap(p)$, where $Cap(p)$ is the capacity of $p$, i.e. the number of points that can be matched to $p$. Now, we compute $C(b_i,k)$ for all $1\leq k\leq Cap(b_i)$ as follows. There exist two steps:
\begin{itemize}
\item [Step 1.]We first check that whether matching the points $\{p \in S\cup T|p\leq b_i\}$ to $b_i$ decreases the cost of the matching or not, as Step 1 of the OMMD algorithm, until $deg(b_i)=Cap(b_i)$:
  \begin{itemize}
    \item If $i=1$, we must determine whether matching the points $a_s, \dots, a_1$ to $b_i$ decreases the cost of the OMMDC or not.
    \item Let $\{a'_1, a'_2, \dots,a'_{s'}\}$ be the set of points matched to $b_{i-1}$. We also check that whether matching the points $a'_1, a'_2, \dots,a'_{s'}$ to $b_i$ decreases the cost of the OMMDC.
     \item If $deg(b_{i-1})=Cap(b_{i-1})$, starting from the smallest point that has been matched to $b_{i-1}$, i.e. $M(b_{i-1},Cap(b_{i-1}))$, we should determine whether matching the points $p\leq M(b_{i-1},Cap(b_{i-1}))$ to $b_i$ decreases the cost of the OMMDC or not (see Figure \ref{fig:10}). This case follows from the limited capacities of the points $b_1,b_2,\dots,b_{i-1}$ (see the OLCMM algorithm given in \cite{Rajabi-Alni2}).

      \end{itemize}

\begin{figure}[h]

\vspace{1cm}
\hspace{0.5cm}
\resizebox{1.8\width}{!}{%

  \includegraphics{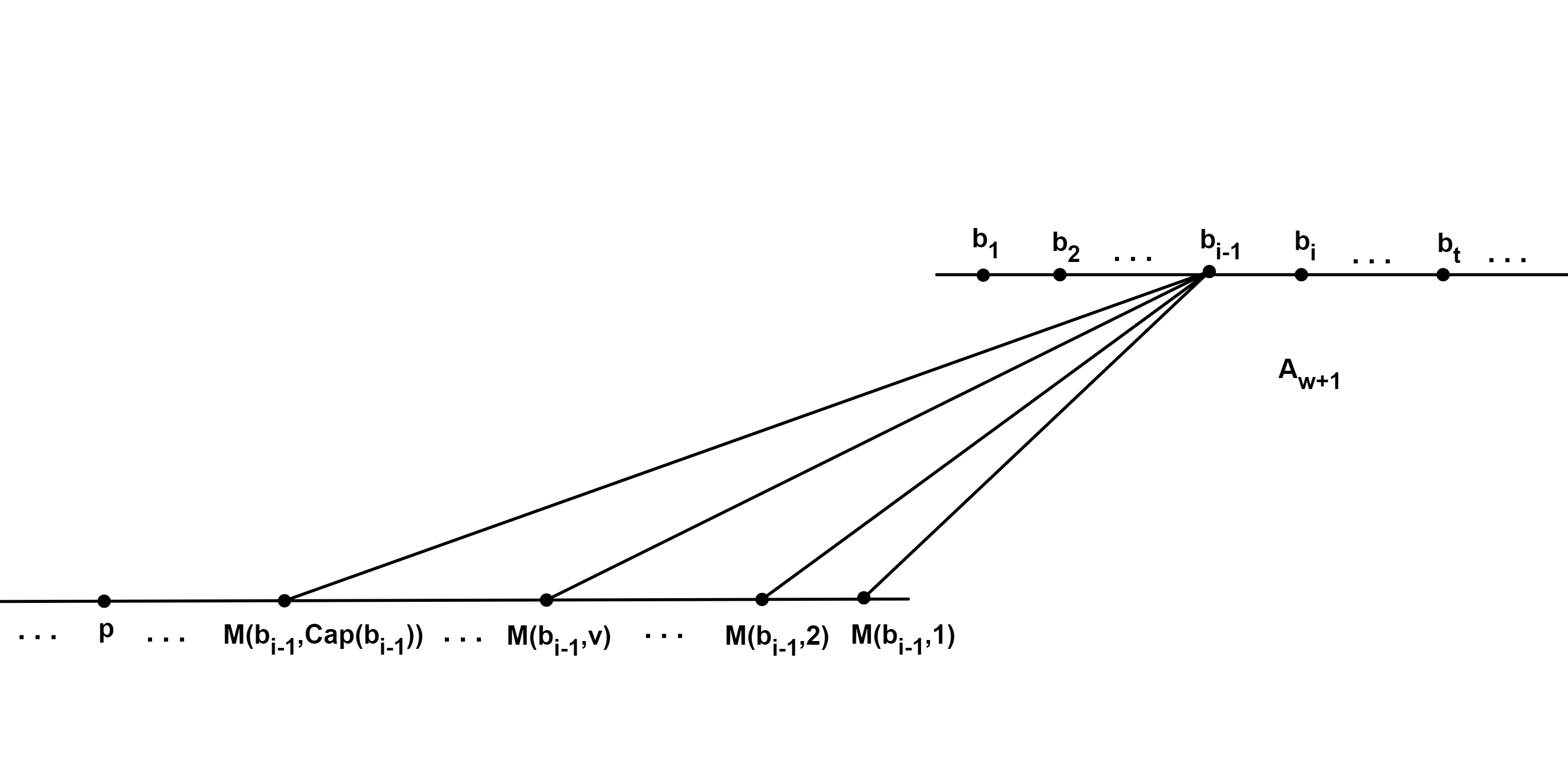}
}
\vspace{-0.5cm}
\caption{In Step 1, we check that whether matching the points $p\leq M(b_{i-1},Cap(b_{i-1}))$ to $b_i$ decreases the cost of the OMMDC.}
\label{fig:10}       
\end{figure}

Let $w'=w+1$. Now, while $deg(b_i)<  Demand(b_i)$ and $w'\geq 1$, two following sub steps run, iteratively:

\item [Step 2.1] In this step, while $deg(b_i)<  Demand(b_i)$, we do as follows:
      \begin{itemize}
        \item If there exists a point $b'_j\in A_{w'}$ with $b'_j\leq b_i$ and $deg(b'_j)> Demand(b'_j)$, we remove the pair $(p',b'_j)$ from the OMMDC and add the pair $(p',b_i)$, where $p'=M(b'_j,deg(b'_j))$.
         \item Otherwise, if there exists a point $b'_j$ in $A_{w'}$ that has been matched to a point $a'_h$ with $Demand(a'_h)<deg(a'_h)<Cap(a'_h)$, by Claim \ref{cccc}, we must match $b_i$ to $a'_h$.
\end{itemize}

If still $deg(b_i)<  Demand(b_i)$, we let $w'=w'-1$ and go to Step 2.2.

\item [Step 2.2] This step is performed exactly as Step 2.2 of the OMMD algorithm.

\end{itemize}
\qed

\section{Conclusions and Future Research}

In this paper, we proposed an $O(n^2)$-time algorithm for computing a one-dimensional many-to-many matching with demands (OMMD). This problem involves finding a many-to-many matching between two point sets on a line where each point is matched to at least a given number of points of the other set. We also briefly described an $O(n^2)$-time algorithm for a generalized version of the OMMD problem, the OMMDC problem, where points have limited capacities in addition to demands. To the best of our knowledge, these are the first $O(n^2)$-time algorithms proposed for both problems. Future work will focus on extending these results to the two-dimensional case and developing new algorithms for practical variants of the OMMD and OMMDC problems that model real-world constraints.


\FloatBarrier

\bibliographystyle{tfs}


\end{document}